\documentclass{JAC2000}
\usepackage{graphicx}
\newcommand{\ud}{\mathrm{d}}
\begin{document}
\title{LOCAL ANALYSIS OF NONLINEAR RMS ENVELOPE DYNAMICS}
\author{A. Fedorova, M. Zeitlin, IPME, RAS, St.~Petersburg, 
V.O. Bolshoj pr., 61, 199178, Russia
\thanks{e-mail: zeitlin@math.ipme.ru}\thanks{http://www.ipme.ru/zeitlin.html;
http://www.ipme.nw.ru/zeitlin.html}}

\maketitle

\begin{abstract}
We present applications of variational -- wavelet approach to 
nonlinear (rational)
rms envelope dynamics.
We have the solution as
a multiresolution (multiscales) expansion in the base of compactly
supported wavelet basis.
\end{abstract}

\section{INTRODUCTION}
In this paper we consider the applications of a new nu\-me\-ri\-cal\--analytical 
technique which is based on the methods of local nonlinear harmonic
analysis or wavelet analysis to the nonlinear 
root-mean-square (rms) envelope dynamics [1].
Such approach may be useful in all models in which  it is 
possible and reasonable to reduce all complicated problems related with 
statistical distributions to the problems described 
by systems of nonlinear ordinary/partial differential 
equations. In this paper we  consider an approach based on 
the second moments of the distribution functions for  the calculation
of evolution of rms envelope of a beam.
The rms envelope equations are the most useful for analysis of the 
beam self--forces (space--charge) effects and also 
allow to consider  both transverse and longitudinal dynamics of
space-charge-dominated relativistic high--brightness
axisymmetric/asymmetric beams, which under short laser pulse--driven
radio-frequency photoinjectors have fast transition from nonrelativistic
to relativistic regime [2]. From the formal point of 
view we may consider rms envelope equations 
after straightforward transformations to standard Cauchy form 
as a system of nonlinear differential equations 
which are not more than rational (in dynamical variables). 
Because of rational type of
nonlinearities we need to consider some extension of our results from 
[3]-[10], which are based on application of wavelet analysis technique to 
variational formulation of initial nonlinear problems.
  
Wavelet analysis is a relatively novel set of mathematical
methods, which gives us a possibility to work with well-localized bases in
functional spaces and give for the general type of operators (differential,
integral, pseudodifferential) in such bases the maximum sparse forms. 
Our approach in this paper is based on the generalization [11] of 
variational-wavelet approach from [3]-[10],
which allows us to consider not only polynomial but rational type of 
nonlinearities.

In part 2 we describe the different forms of rms equations.
In part 3 we present explicit analytical construction for solutions of
rms equations from part 2, which are based on our variational formulation 
of initial dynamical problems and on  multiresolution representation [11].
We give explicit representation for all dynamical variables in the base of
compactly supported wavelets. Our solutions
are parametrized
by solutions of a number of reduced algebraical problems from which one
is nonlinear with the same degree of nonlinearity and the rest  are
the linear problems which correspond to particular
method of calculation of scalar products of functions from wavelet bases
and their derivatives. 
In part 4 we consider results of numerical calculations.

\section{RMS EQUATIONS}
Below we consider a number of different forms of RMS envelope equations,
which are from the formal  point of view
not more than nonlinear differential equations with rational
nonlinearities and variable coefficients.
Let $f(x_1,x_2)$ be the distribution function which gives full information
 about 
noninteracting ensemble of beam particles regarding to trace space or 
transverse phase coordinates $(x_1,x_2)$. 
Then we may extract the first nontrivial bit of `dynamical information' from 
the second moments
\begin{eqnarray}
\sigma_{x_1}^2&=&<x_1^2>=\int\int x_1^2 f(x_1,x_2)\ud x_1\ud x_2 \nonumber\\
\sigma_{x_2}^2&=&<x_2^2>=\int\int x_2^2 f(x_1,x_2)\ud x_1\ud x_2 \\
\sigma_{x_1 x_2}^2&=&<x_1 x_2>=\int\int x_1 x_2 f(x_1,x_2)\ud x_1\ud x_2 \nonumber
\end{eqnarray}
RMS emittance ellipse is given by
$
\varepsilon^2_{x,rms}=<x_1^2><x_2^2>-<x_1 x_2>^2
$.
 Expressions for twiss  parameters are also based on the second moments.

We will consider the following particular
cases of rms envelope equations, which described evolution
of the moments (1) ([1],[2] for full designation):
for asymmetric beams we have the system of two envelope equations
of the second order for $\sigma_{x_1}$ and $\sigma_{x_2}$:
\begin{eqnarray}
&&\sigma^{''}_{x_1}+\sigma^{'}_{x_1}\frac{\gamma '}{\gamma}+
\Omega^2_{x_1}\left(\frac{\gamma '}{\gamma}\right)^2\sigma_{x_1}=\\
&&{I}/({I_0(\sigma_{x_1}+\sigma_{x_2})\gamma^3})
+
\varepsilon^2_{nx_1}/{\sigma_{x_1}^3\gamma^2},\nonumber\\
&&\sigma^{''}_{x_2}+\sigma^{'}_{x_2}\frac{\gamma '}{\gamma}+
\Omega^2_{x_2}\left(\frac{\gamma '}{\gamma}\right)^2\sigma_{x_2}=\nonumber\\ 
&&{I}/({I_0(\sigma_{x_1}+\sigma_{x_2})\gamma^3})
+
\varepsilon^2_{nx_2}/{\sigma_{x_2}^3\gamma^2}\nonumber
\end{eqnarray}
The envelope equation for an axisymmetric beam is
a particular case of preceding equations.

Also we have related Lawson's equation for evolution of the rms
envelope in the paraxial limit, which governs evolution of cylindrical
symmetric envelope under external linear focusing channel
of strenghts $K_r$:
\begin{equation}
\sigma^{''}+\sigma^{'}\left(\frac{\gamma '}{\beta^2\gamma}\right)+
K_r\sigma=\frac{k_s}{\sigma\beta^3\gamma^3}+
\frac{\varepsilon^2_n}{\sigma^3\beta^2\gamma^2},\nonumber
\end{equation}
where
$
K_r\equiv -F_r/r\beta^2\gamma mc^2, \ \ \
 \beta\equiv \nu_b/c=\sqrt{1-\gamma^{-2}}
$

After transformations to Cauchy form we can see that
all this equations from the formal point of view are not more than
ordinary differential equations with rational nonlinearities
and variable coefficients
(also,we may consider regimes in which $\gamma$, $\gamma'$
are not fixed functions/constants but satisfy some additional differential 
constraint/equation,
but this case does not change our general approach).

\section{Rational Dynamics}

The first main part of our consideration is some variational approach
to this problem, which reduces initial problem to the problem of
solution of functional equations at the first stage and some
algebraical problems at the second stage.
We have the solution in a compactly
supported wavelet basis.
An example of such type of basis is demonstrated on Fig.~1.
\begin{figure}[htb]
\centering
\includegraphics*[width=40mm]{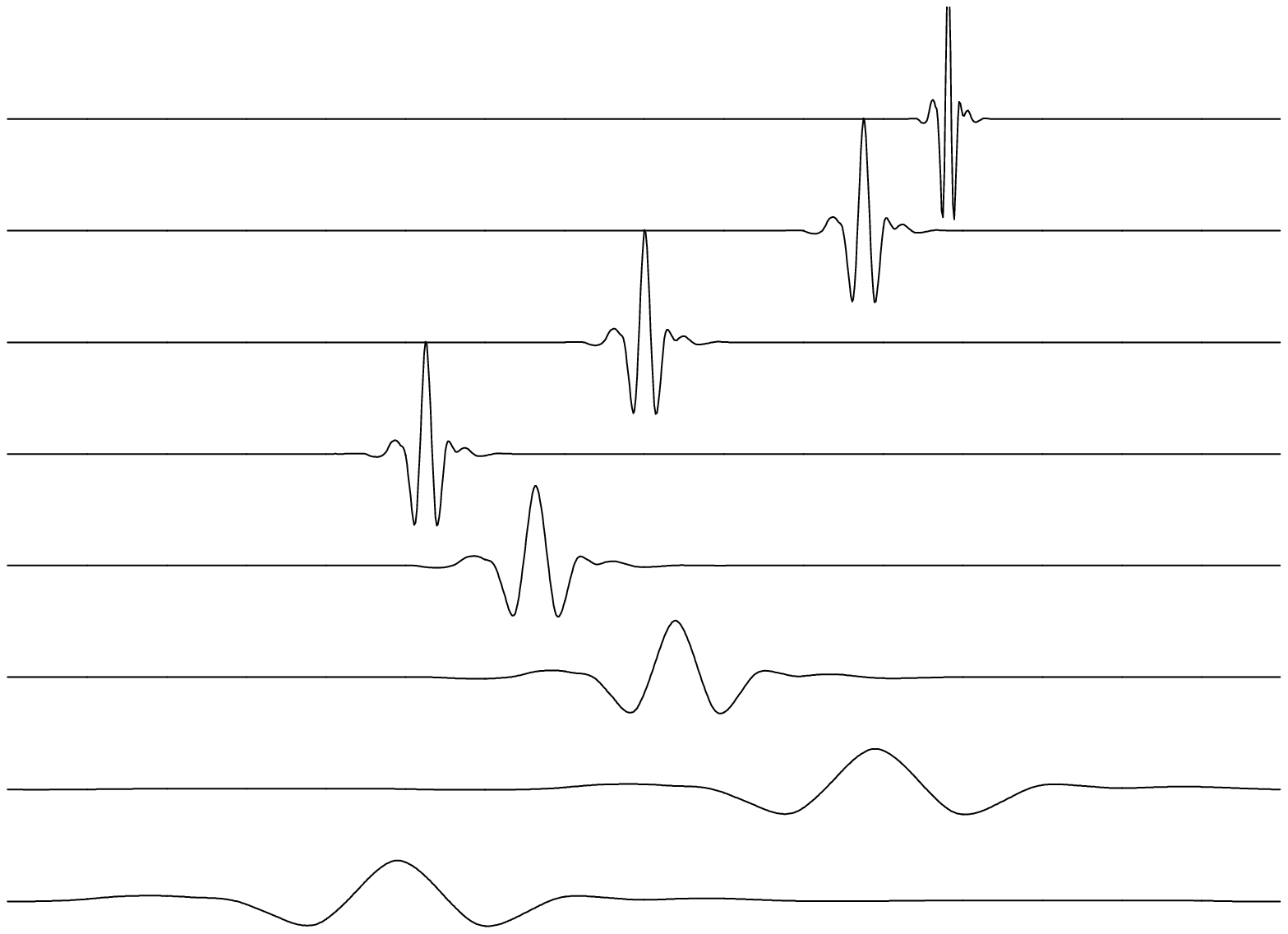}
\caption{Wavelets at different scales and locations.}
\end{figure}
Multiresolution representation is the second main part of our construction.
The solution is parameterized by solutions of two reduced algebraical
problems, one is nonlinear and the second are some linear
problems, which are obtained from one of the standard wavelet
constructions:  the method of Connection
Coefficients (CC) or  Stationary Subdivision Schemes (SSS).

So, our variational-multiresolution approach [11] gives us 
possibility to construct explicit numerical-analytical
solution for the following systems of nonlinear differential 
equations 
\begin{equation}
\dot{z}=R(z,t) \quad \mbox{or} \quad Q(z,t)\dot{z}=P(z,t),
\end{equation}
where $z(t)=(z_1(t),...,z_n(t))$ is the vector of dynamical variables 
$z_i(t)$,

$R(z,t)$ is not more than rational function of z,

$P(z,t), Q(z,t)$ are not more than polynomial functions of z and 
P,Q,R have arbitrary dependence of time. 

The solution has the following form
\begin{equation}\label{eq:z}
z(t)=z_N^{slow}(t)+\sum_{j\geq N}z_j(\omega_jt), \quad \omega_j\sim 2^j
\end{equation}
which corresponds to the full multiresolution expansion in all time 
scales.
Formula (\ref{eq:z}) gives us expansion into a slow part $z_N^{slow}$
and fast oscillating parts for arbitrary N. So, we may move
from coarse scales of resolution to the 
finest one for obtaining more detailed information about our dynamical process.
The first term in the RHS of representation (5) corresponds on the global level
of function space decomposition to  resolution space and the second one
to detail space. In this way we give contribution to our full solution
from each scale of resolution or each time scale.
The same is correct for the contribution to power spectral density
(energy spectrum): we can take into account contributions from each
level/scale of resolution.

So, we have the solution of the initial nonlinear
(rational) problem  in the form
\begin{eqnarray}\label{eq:pol3}
z_i(t)=z_i(0)+\sum_{k=1}^N\lambda_i^k Z_k(t),
\end{eqnarray}
where coefficients $\lambda_i^k$ are roots of the corresponding
reduced algebraical (polynomial) problem [11].
Consequently, we have a parametrization of solution of initial problem
by solution of reduced algebraical problem.

So, the obtained solutions are given
in the form (\ref{eq:pol3}),
where
$Z_k(t)$ are basis functions and
  $\lambda_k^i$ are roots of reduced
 system of equations.  In our  case $Z_k(t)$
are obtained via multiresolution expansions and represented by
 compactly supported wavelets and $\lambda_k^i$ are the roots of
reduced polynomial system with coefficients, which
are given by CC or SSS  constructions.

Each $Z_j(t)$ is a representative of corresponding multiresolution 
subspace $V_j$,
which is a member of the sequence of increasing closed subspaces $V_j$:
\begin{equation}
...V_{-2}\subset V_{-1}\subset V_0\subset V_{1}\subset V_{2}\subset ...
\end{equation}

The basis in each $V_j$ is 
\begin{equation}
\varphi_{jl}(x)=2^{j/2}\varphi(2^j x-\ell)
\end{equation}
where indices $\ell, j$ represent translations and scaling 
respectively or action of underlying affine group
which act as a ``microscope'' and allow us to construct
corresponding solution with needed level of resolution.

It should be noted that such representations (5),(6)
for solutions of equations (2),(3) give the best possible localization
properties in corresponding phase space.This is especially important because 
our dynamical variables corresponds to moments of ensemble of beam particles.

\section{Numerical Calculations}

In this part we consider numerical illustrations of previous analytical
approach. Our numerical calculations are based on compactly supported
Daubechies wavelets and related wavelet families.
On Fig.~2 we present according to formulae (5),(6) contributions 
to approximation of our dynamical evolution (top row on the Fig.~3) 
starting from
the coarse approximation, corresponding to scale $2^0$ (bottom row)
to the finest one corresponding to the scales from $2^1$ to  $2^5$
or from slow to fast components (5 frequencies) as details for approximation.
Then on Fig.~3, from bottom to top, we demonstrate the summation
of contributions from corresponding levels of resolution given on
Fig.~2 and as result we restore via 5 scales (frequencies) approximation
our dynamical process(top row on Fig.~3 ).

We also produce the same decomposition/approximation on 
the level of power spectral density (Fig.~4).
It should be noted that complexity of such algorithms are minimal regarding
other possible.
Of course, we may use different multiresolution analysis schemes, which
are based on different families of generating wavelets and apply
such schemes of numerical--analytical calculations to any dynamical
process which may be described by systems of ordinary/partial differential
equations with rational nonlinearities [11].
 
We would like to thank Professor
James B. Rosenzweig and Mrs. Melinda Laraneta for
nice hospitality, help and support during
UCLA ICFA Workshop.
\begin{figure}
\centering
\includegraphics*[width=60mm]{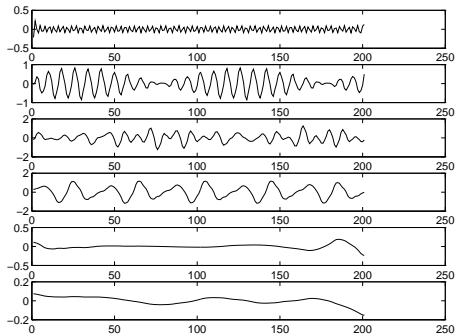}
\caption{Contributions to approximation: from scale $2^1$ to $2^5$.}
\end{figure}
\begin{figure}
\centering
\includegraphics*[width=60mm]{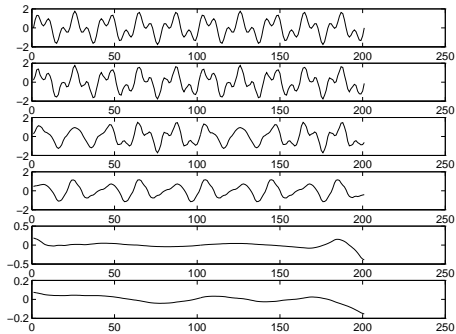}
\caption{Approximations: from scale $2^1$ to $2^5$.}
\end{figure}
\begin{figure}
\centering
\includegraphics*[width=60mm]{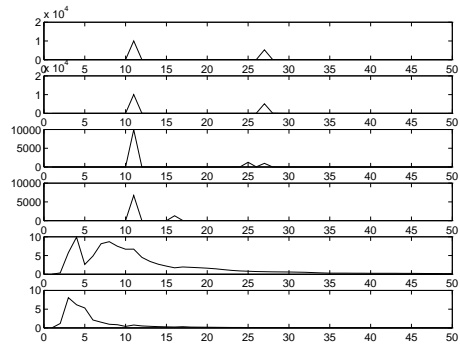}
\caption{Power spectral density:  from scale $2^1$ to $2^5$.}
\end{figure}

\end{document}